\documentclass{article}
\pdfoutput=1
\usepackage{jcappub}

\usepackage{amsmath}
\usepackage{enumitem}
\usepackage{graphicx}

\usepackage[caption=false]{subfig}

\usepackage[range-phrase=\textendash, range-units=single]{siunitx}

\let\sun\odot
\DeclareSIUnit\solarmass{\ensuremath{M_\sun}}
\DeclareSIUnit\erg{erg}
\DeclareSIUnit\parsec{pc}

\renewcommand{\vec}[1]{\boldsymbol{#1}}

\subheader{Published in JCAP}
\title{Search for neutrino counterparts to the gravitational wave sources from LIGO/Virgo O3 run with the ANTARES detector}

\author[a,b]{A.~Albert}
\author[c]{S.~Alves}
\author[d]{M.~Andr\'e}
\author[e]{M.~Ardid}
\author[e]{S.~Ardid}
\author[f]{J.-J.~Aubert}
\author[g]{J.~Aublin}
\author[g]{B.~Baret}
\author[h]{S.~Basa}
\author[g]{Y.~Becherini}
% ...

\affiliation[a]{Universit\'e de Strasbourg, CNRS,  IPHC UMR 7178, F-67000 Strasbourg, France} 
\affiliation[b]{ Universit\'e de Haute Alsace, F-68100 Mulhouse, France} 
\affiliation[c]{IFIC - Instituto de F\'isica Corpuscular (CSIC - Universitat de Val\`encia) c/ Catedr\'atico Jos\'e Beltr\'an, 2 E-46980 Paterna, Valencia, Spain} 
\affiliation[d]{Technical University of Catalonia, Laboratory of Applied Bioacoustics, Rambla Exposici\'o, 08800 Vilanova i la Geltr\'u, Barcelona, Spain} 
\affiliation[e]{Institut d'Investigaci\'o per a la Gesti\'o Integrada de les Zones Costaneres (IGIC) - Universitat Polit\`ecnica de Val\`encia. C/  Paranimf 1, 46730 Gandia, Spain} 
\affiliation[f]{Aix Marseille Univ, CNRS/IN2P3, CPPM, Marseille, France} 
\affiliation[g]{Universit\'e Paris Cit\'e, CNRS, Astroparticule et Cosmologie, F-75013 Paris, France} 
\affiliation[h]{Aix Marseille Univ, CNRS, CNES, LAM, Marseille, France } 
% ...

\collaboration{The ANTARES Collaboration}

\emailAdd{antares.spokesperson@in2p3.fr}

\abstract{Since 2015 the LIGO and Virgo interferometers have detected gravitational waves from almost one hundred coalescences of compact objects (black holes and neutron stars). This article presents the results of a search performed with data from the ANTARES telescope to identify neutrino counterparts to the gravitational wave sources detected during the third LIGO/Virgo observing run and reported in the catalogues GWTC-2, GWTC-2.1, and GWTC-3. This search is sensitive to all-sky neutrinos of all flavours and of energies $> \SI{100}{\giga\electronvolt}$, thanks to the inclusion of both track-like events (mainly induced by $\nu_\mu$ charged-current interactions) and shower-like events (induced by other interaction types). Neutrinos are selected if they are detected within $\pm \SI{500}{\second}$ from the GW merger and with a reconstructed direction compatible with its sky localisation. No significant excess is found for any of the 80 analysed GW events, and upper limits on the neutrino emission are derived. Using the information from the GW catalogues and assuming isotropic emission, upper limits on the total energy $E_{\rm tot, \nu}$ emitted as neutrinos of all flavours and on the ratio $f_\nu = E_{\rm tot, \nu}/E_{\rm GW}$ between neutrino and GW emissions are also computed. Finally, a stacked analysis of all the 72 binary black hole mergers (respectively the 7 neutron star - black hole merger candidates) has been performed to constrain the typical neutrino emission within this population, leading to the limits: $E_{\rm tot, \nu} < \SI{4.0e53}{\erg}$ and $f_\nu < 0.15$ (respectively, $E_{\rm tot, \nu} < \SI{3.2e53}{\erg}$ and $f_\nu < 0.88$) for $E^{-2}$ spectrum and isotropic emission. Other assumptions including softer spectra and non-isotropic scenarios have also been tested.}

\arxivnumber{2302.07723}

\begin{document}

\maketitle

\section{Introduction}
\label{sec:intro}

Since the first detection of gravitational waves (GWs) from compact binary mergers in 2015~\cite{LIGOScientific:2016aoc}, GW interferometers have opened a new window on the Universe, complementary to the ones already being explored with other cosmic messengers (cosmic rays, photons, neutrinos). These capabilities have already allowed the association of the GW signal GW170817 emitted by the merger of a binary neutron star system with the emission of a short gamma-ray burst (GRB), GRB 170817A, detected by Fermi and INTEGRAL in gamma rays, and the related afterglow across a wider range of the electromagnetic spectrum~\citep{LIGOScientific:2017ync}. 

Neutrinos are also expected to be emitted from the relativistic outflows that characterise such mergers: see e.g., \cite{Kimura:2018vvz} for binary neutron star (BNS) mergers, \cite{Kimura:2017kan} for neutron star-black hole (NSBH) mergers, and \cite{Kotera:2016dmp} for binary black hole (BBH) mergers. Previous searches performed with ANTARES \citep{ANTARES:2020kxp}, Baikal-GVD \citep{Baikal-GVD:2018cya}, IceCube \citep{Abbasi:2022rbd}, and Super-Kamiokande \citep{Super-Kamiokande:2021dav} have not been able to identify an excess of neutrinos and upper limits have been reported.

This article presents an updated search using the latest GW catalogues covering detections in 2019-2020 and the ANTARES data from the same period. Besides the follow-up of individual events, performed in the same way as in previous publications, first population studies are also presented by carrying out a stacking analysis for binary mergers of the same nature and taking advantage of the extensive catalogue reported by the GW community.

\subsection{The GW catalogues}
\label{sec:gwtc}

This paper focuses on GW sources detected during the third observing run (O3) of the LIGO and Virgo detectors and reported in the following official LIGO/Virgo catalogues:
\begin{itemize}
    \item GWTC-2 \citep{LIGOScientific:2020ibl}: this catalogue reports detections made during the first half of O3 (April - September 2019). It contains 39 candidates, including 1 BNS, 2 NSBH, and 36 BBH events.
    \item GWTC-2.1 \citep{LIGOScientific:2021usb}: this is an update of GWTC-2, with 8 additional events not reported in the previous catalogue and that have a high probability of astrophysical origin. The catalogue also includes a larger selection of $\sim 1.2$\,k events with a false alarm rate of less than $2$ per day but lower astrophysical probability, which are not considered in this analysis.
    \item GWTC-3 \citep{LIGOScientific:2021djp}: this catalogue covers the second half of O3 (November 2019 - March 2020) and contains 35 objects, including 4 NSBH candidates. Seven marginal candidates that do not fully satisfy the criteria for an astrophysical origin are also reported, of which only GW200105\_162426 is kept as it has also been reported independently as a plausible NSBH candidate \citep{LIGOScientific:2021qlt}. 
\end{itemize}

For each of these objects, the LIGO-Virgo collaboration provides a FITS file~\citep{FITS} with the timing of the merger $t_{\rm GW}$ and the constraints on the source direction $\Omega$ as a skymap $\mathcal{P}(\Omega)$, as well as posterior samples containing all the correlations between source direction $\Omega$, luminosity distance estimate $D_L$, and other source parameters such as the masses of the two merging objects $m_{1,2}$ (with the convention $m_1 > m_2$), the energy radiated in gravitational waves $E_{\rm GW}$ (defined as the difference between the estimated mass of the final object and the sum of the masses of the initial objects), and the inclination between the total angular momentum and the line-of-sight $\theta_{\rm jn}$. The classification among the different categories is made based on the mass estimates: BNS if $m_2 < m_1 < \SI{3}{\solarmass}$, NSBH if $m_2 < \SI{3}{\solarmass} < m_1$, BBH otherwise. When considering the different catalogues, 83 objects are selected, including 1 BNS and 7 NSBH candidates.

\subsection{The ANTARES telescope}
\label{sec:antares}

The ANTARES neutrino telescope~\citep{ANTARES:2011hfw}, located in the depths of the Mediterranean Sea, offshore from Toulon (France), has been operating in its final configuration between May 2008 and February 2022. It was composed of an array of 885 photomultiplier tubes (PMTs) enclosed in pressure-proof glass spheres, arranged in triplets over 12 vertical lines, spaced by $\sim \SI{70}{\meter}$ and anchored at a depth of $\sim \SI{2475}{\meter}$. 

The PMTs detect the Cherenkov light induced by relativistic charged particles originating in the interaction of a neutrino with matter surrounding the detector; the space and time pattern of PMT signals (or hits) allows the neutrino properties (direction and energy) to be inferred. Charged-current interactions of muon (anti-)neutrinos are characterised by the presence of a long muon track; the typical angular resolution for such events is $<\SI{1}{\degree}$ for $E_\nu >\SI{100}{\giga\electronvolt}$ \citep{ANTARES:2011vtx}. Other types of neutrino interactions only produce hadronic (and, in the case of $\nu_e$ charged-current interactions, electromagnetic) showers, with a more localised light deposit and compact topology. Despite the shorter lever arm with respect to the muon tracks, ANTARES still achieved a median angular resolution of a few degrees for the neutrino direction \citep{ANTARES:2017ivh}. In the following, the two event categories described above are referred to as the track and shower samples respectively.

ANTARES data were organised in consecutive runs of at most twelve hours. The present analysis uses data from 2019 and 2020 to identify neutrino counterparts to the GW emissions described in \autoref{sec:gwtc}. A larger dataset including data from January 2018 to December 2020 is used for background estimation.

Several searches for neutrino counterparts to GW events with ANTARES have been carried out in the past. These studies were limited to event-by-event follow-ups and reported null results. See \cite{ANTARES:2017fqy} for GW170104, \cite{ANTARES:2017bia} for GW170817, and \cite{ANTARES:2020kxp} for other O2 events.

\section{Analysis method}
\label{sec:analysis}

This search focuses on the selection of neutrino events in a time window of \SI{1000}{\second} centred on the GW emission time $t_{\rm GW}$, as motivated in \citep{Baret:2011tk}, and in the region $\mathcal{R}_{90}$ containing 90\% of the source localisation probability, as built directly from the GW 2D skymaps $\mathcal{P}(\Omega)$. As the reconstructed event direction does not match the true neutrino direction perfectly due to the scattering angle and the finite detector resolution, this region of interest (RoI) is extended by an angle $\alpha$ to account for these effects, meaning that an event with direction $\vec{x}$ would be selected if $\min_{\vec{d} \in \mathcal{R}_{90}}(\arccos{\left(\vec{x} \cdot \vec{d} \right)}) \leq \alpha$. This extended angle $\alpha$ is a free parameter of the analysis, whose choice is based on the optimisation method described below.

The ANTARES events are divided into four categories according to whether they are classified as tracks or showers and whether their reconstructed direction is upgoing or downgoing, each case based on a specific selection and optimisation procedure, as described in sections~\ref{sec:tracks} and~\ref{sec:showers}.

The reconstructed data are largely dominated by atmospheric muons. A selection is then applied to reduce it to an expected number of events $B=\num{2.7e-3}$, such that the detection of one event would correspond to a $3\sigma$ excess ($3\sigma$ condition). This condition is separately set for each category. For a given GW, the background expectation is estimated using the dataset from 2018 to 2020. Only runs with similar data-taking conditions as the ones during the ANTARES run $r_{\rm GW}$ overlapping with the GW time, characterised by the mean burst fraction\footnote{This quantity is defined by estimating how often, in a given run, the PMT counting rate is more than 20\% higher than the baseline value for this run. This quantity has been found to be correlated with the detector noise level, whose evolution is mainly driven by deep-sea bioluminescent emissions \citep{10.1371/journal.pone.0067523}.}, are selected. As illustrated in the left panel of \autoref{fig:ana:tracks}, this procedure is found to allow for a proper estimate of the background while ensuring a better characterisation of the tails of the distribution as compared to a statistically-limited estimation using only the data from the run $r_{\rm GW}$.

A dedicated Monte Carlo (MC) simulation~\cite{ANTARES:2020bhr} has been produced to be used in this optimisation process. Simulations are done on a run-by-run basis, where each run has a specific simulation to reproduce the particular environmental and detector conditions during this run. For GW events with precise sky localisation ($\mathcal{R}_{90}$ region smaller than $\SI{3000}{\square\deg}$), additional neutrinos generated solely within this region have also been produced to accurately estimate the corresponding detector acceptance for an $E^{-\gamma}$ spectrum ($= \int A_{\rm eff}(E) E^{-\gamma} {\rm d}E$, where $A_{\rm eff}(E)$ is the effective area). More details about the detector acceptance for a given neutrino spectrum are in \cite{ANTARES:2017dda}.

The final cuts are optimised to ensure that the expected number of selected background events after all cuts is fulfilling the $3\sigma$ condition defined above.

\subsection{Track event selection}
\label{sec:tracks}

The track selection procedure is adapted from the one presented in \cite{ANTARES:2017fqy}. The upgoing and downgoing (respectively with reconstructed incoming direction below or above the horizon) event selections are different, as the latter is more likely to be contaminated by the atmospheric muon background and needs extra care.

\begin{figure}[hbtp]
    \subfloat[Estimation of the background.]{\includegraphics[width=0.48\textwidth]{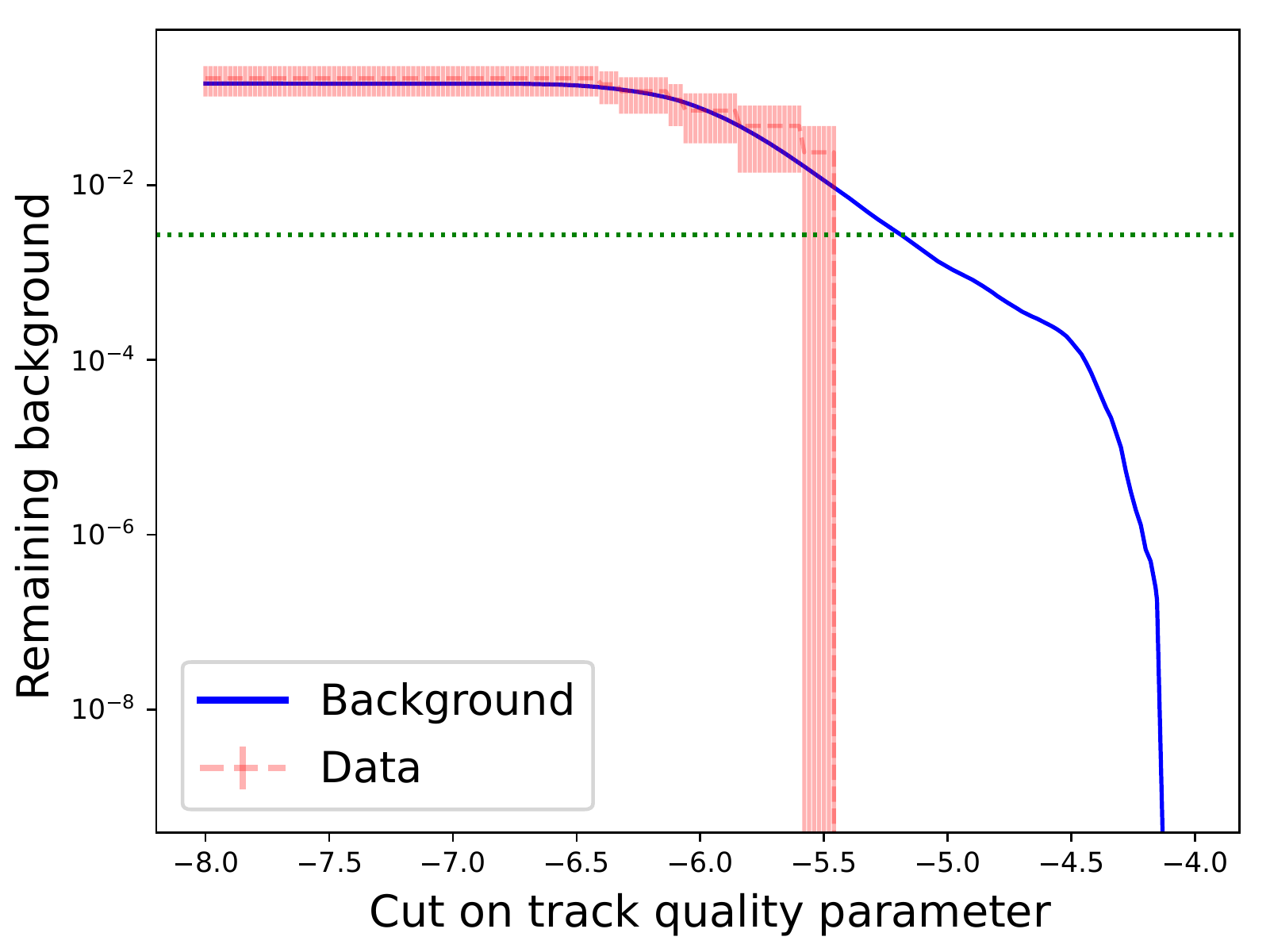}}
    \subfloat[Optimisation of the cuts.]{\includegraphics[width=0.48\textwidth]{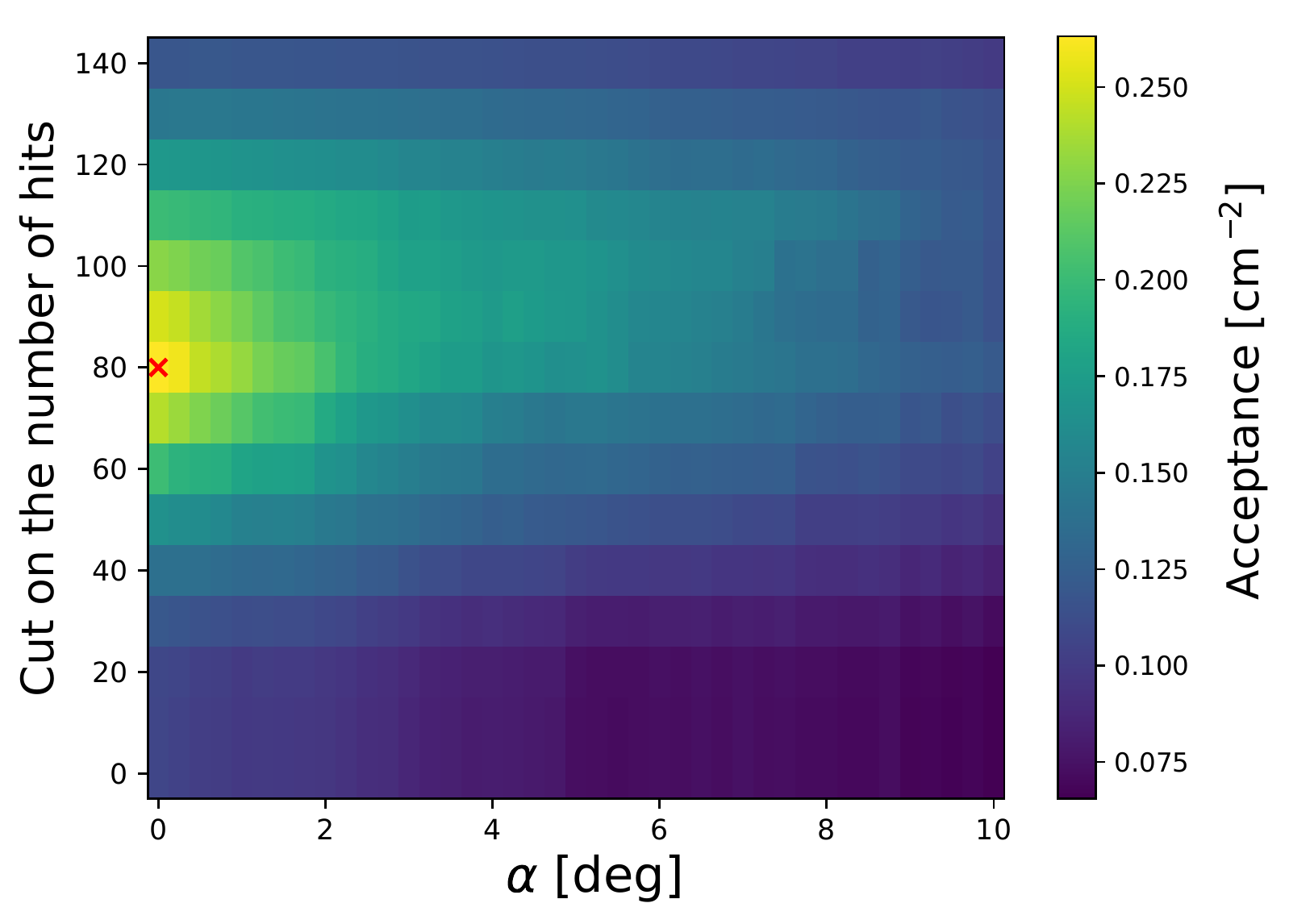}}
    \caption{Illustration of the analysis procedure for the downgoing track category for GW190521. The left panel shows the expected remaining background after varying cuts on the track quality parameter $\Lambda$: the blue curve shows the estimation based on the procedure described in the previous paragraphs, the red histogram shows the background estimated using only the ANTARES 12-hour run containing the signal time window with its statistical uncertainty, and the dotted line is the required background to ensure the $3\sigma$ condition. The right panel shows the average acceptance, assuming an $E^{-2}$ neutrino spectrum, as a function of the cut on the number of hits and the value of $\alpha$. For each bin, the cut on $\Lambda$ is fixed to the value that satisfies exactly the $3\sigma$ condition. The optimal working point, maximising the acceptance, is indicated by the red cross.}
    \label{fig:ana:tracks}
\end{figure}

For upgoing tracks, a cut on the track reconstruction quality parameter, $\Lambda$ \cite{ANTARES:2013tra}, is applied. The values of $\alpha$ and of the cut on $\Lambda$ are optimised by ensuring the $3\sigma$ condition described above, as well as maximising the signal acceptance in the hypothesis of an $E^{-2}$ spectrum. For downgoing events, the procedure is similar except that an additional cut on the number of hits employed for the reconstruction is also applied, as illustrated in the right panel of \autoref{fig:ana:tracks}.

\subsection{Shower event selection}
\label{sec:showers}
 
The selection steps for neutrino interactions yielding showering events are similar to the ones presented in \cite{ANTARES:2020kxp}. Events must be contained within the detector and must not be classified as a track by the selection described in the previous section. The discrimination between neutrinos and atmospheric muons is achieved thanks to an extended likelihood ratio $\mathcal{L}_\mu$ defined by comparing the neutrino and muon hypotheses for each hit associated with the shower on the basis of its deposited charge, timing, and distance to the reconstructed shower position \citep{ANTARES:2017ivh}. Another parameter is used to further reduce the background contamination: for upgoing events, this parameter is defined from a Random Decision Forest (RDF) classifier \citep{Folger:2014rct} while the downgoing selection exploits the number of hits used in the event fitting.

\begin{figure*}[hbtp]
    \subfloat[Estimation of the background.]{\includegraphics[width=0.48\textwidth]{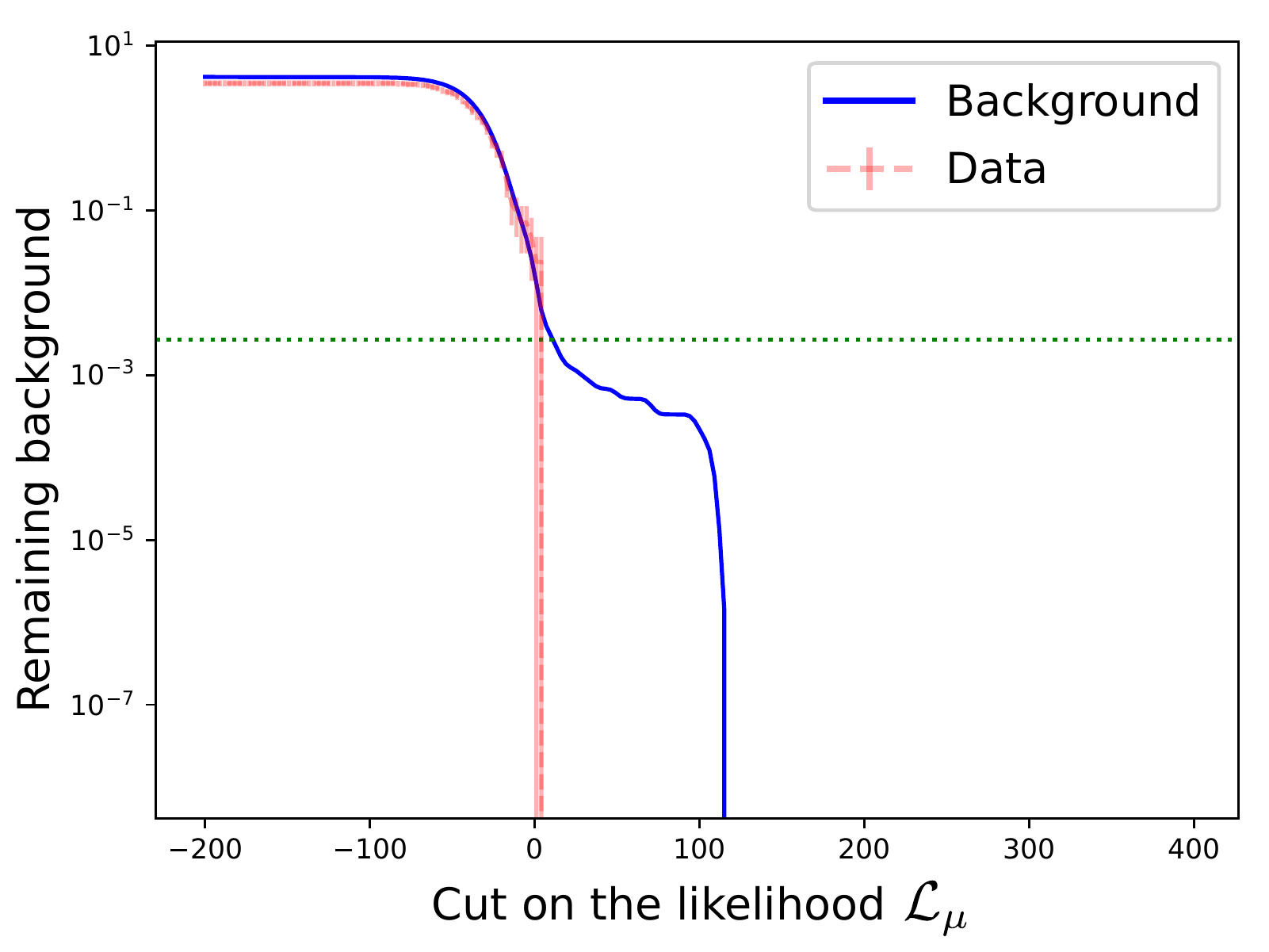}}
    \subfloat[Optimisation of the cuts.]{\includegraphics[width=0.48\textwidth]{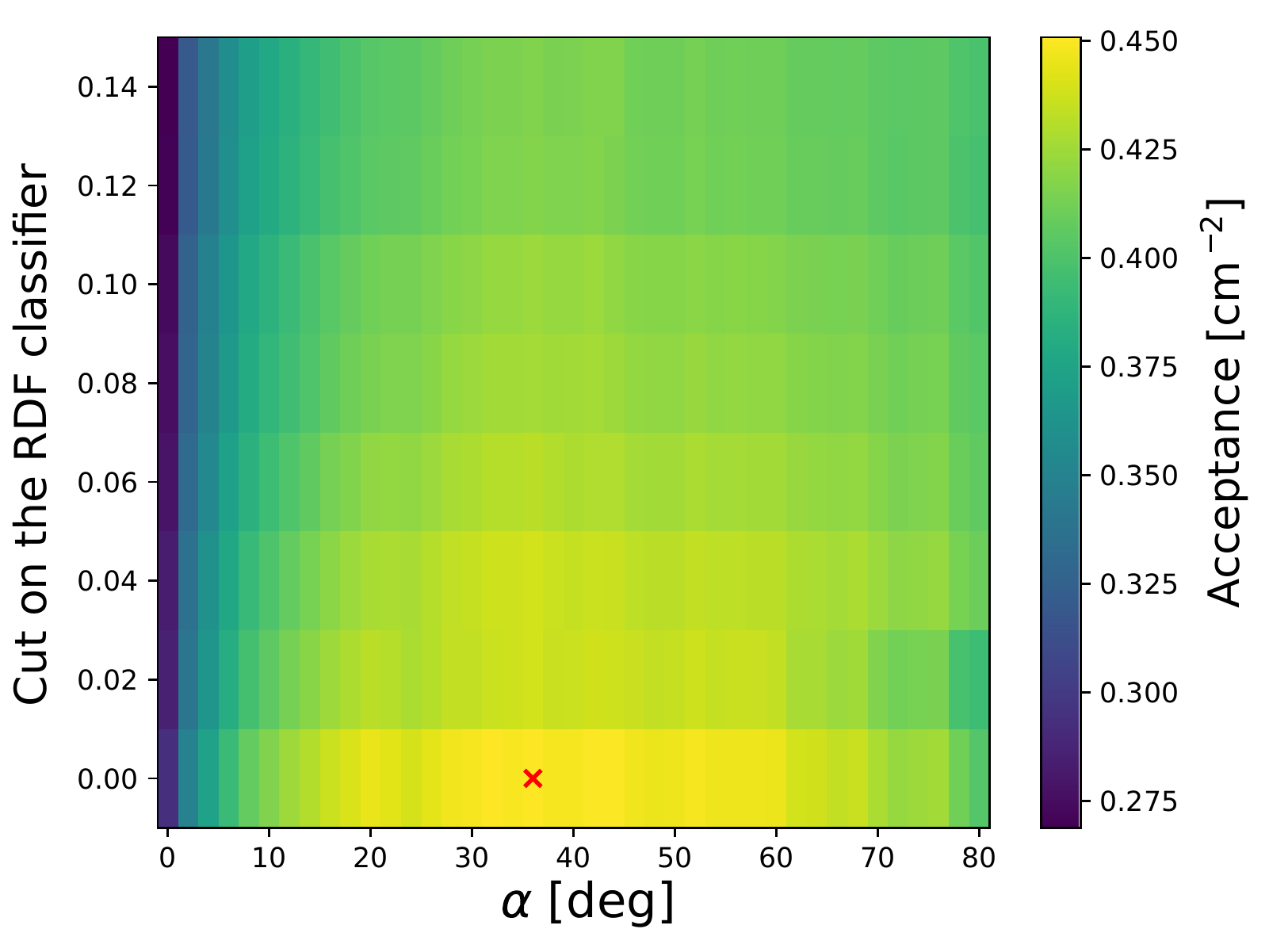}}
    \caption{Illustration of the analysis procedure for the downgoing shower category for GW190521. The left panel shows the expected remaining background after varying cuts on $\mathcal{L}_\mu$ with the same colour code and line style as in \autoref{fig:ana:tracks}. The right panel shows the average acceptance, assuming an $E^{-2}$ neutrino spectrum, as a function of the cut on the RDF classifier and the value of $\alpha$. For each bin, the cut on $\mathcal{L}_\mu$ is fixed to the value that satisfies exactly the $3\sigma$ condition. The optimal working point, maximising the acceptance, is indicated by the red cross.}
    \label{fig:ana:showers}
\end{figure*}

As for the track selection, the values of the cuts on these parameters and on the extension of the RoI $\alpha$ are optimised to ensure the $3\sigma$ condition and to maximise the acceptance, as illustrated in \autoref{fig:ana:showers}.

\subsection{Detector systematics}
\label{sec:syst}

Several systematic effects may affect the detector performance, hence the obtained constraints on the neutrino emission. Three sources of uncertainty, found to be the dominant effects as already described e.g. in \cite{ANTARES:2020kxp}, are taken into account and evaluated independently for the four event categories (track/showers upgoing/downgoing):
\begin{itemize}
\item The first one is related to the uncertainty on the PMT photon detection efficiency and on the water absorption length; the related uncertainties have been re-evaluated by varying these two quantities within a typical interval of  $\pm 10\%$ in dedicated MC simulations and estimating the overall impact on the signal acceptance.
\item The second source is linked to the capability of the run-by-run MC simulations to properly reproduce data conditions; the related error is estimated by comparing the variability of event rates between data and simulations.
\item The last effect is the combined statistical and systematic uncertainty on the background expectation. The related uncertainty is obtained by varying the list of similar runs employed for its estimation.
\end{itemize}

The total uncertainties on the acceptance related to the first two sources are 18\%, 14\%, 21\%, and 19\% respectively for the upgoing tracks, downgoing tracks, upgoing showers, and downgoing showers. The overall uncertainty on the background is about 20\% for all event categories.

\section{Statistical analysis}
\label{sec:stat}

For each GW event, the number of observed neutrino candidates in time and spatial coincidence in each category can be converted into a significance of the observation using Poisson statistics. In the absence of any excess of neutrino events with respect to the background expectation, upper limits on the neutrino emission are calculated. In the following section, several assumptions are made:
\begin{enumerate}[label=(\roman*)]
    \item The source localisation is supposed to be within $\mathcal{R}_{90}$ (as this is the region for which the selection has been optimised). Therefore, the GW posterior samples are restricted to those with $\Omega \in \mathcal{R}_{90}$ and the final constraints neglect the chances for the actual source to be localised in the rest of the sky.
    \item There is equipartition between the neutrino flavours at Earth due to the averaging of oscillations over astronomical distances \citep{Bustamante:2019sdb}, starting from $\nu_e:\nu_\mu:\nu_\tau = 1:2:0$ at production. This allows reporting limits on the all-flavour neutrino emission.
    \item The neutrino energy spectrum is described by a single power law ${\rm d}N/{\rm d}E = \phi \cdot (E/\si{\giga\electronvolt})^{-\gamma}$ where $\phi$ is expressed in units of \si{\per\giga\electronvolt\per\square\centi\meter} and $\gamma$ is the spectral index. The nominal case is $\gamma=2$ ($E^{-2}$ spectrum).
\end{enumerate}

The ANTARES acceptance is estimated using the MC simulations described in \autoref{sec:analysis}. It depends on the shape of the assumed neutrino spectrum (characterised by $\gamma)$, on the source direction $\Omega$, and on the event category $c$. It is averaged over the neutrino flavours and can be decomposed into a normalisation factor and a direction-dependent component: $\mathcal{A}_\gamma^{(c)}(\Omega) = a_\gamma^{(c)} \cdot f_\gamma^{(c)}(\Omega)$.

\subsection{Constraints on the neutrino flux}
\label{sec:flux}

This section presents the limits on the overall flux normalisation $\phi$ obtained for an all-flavour emission with $\gamma=2$. The cut-and-count analysis described in the previous sections corresponds to a Poisson likelihood
\begin{equation}
    \mathcal{L} \left( \{N^{(c)}\}; \{B^{(c)}\}, \{a_\gamma^{(c)}\}, \Omega, \phi, \gamma \right) = \prod_{c \in \mathcal{C}} \textrm{Poisson}\left( N^{(c)}; B^{(c)} + \phi \cdot a_\gamma^{(c)} \cdot f_\gamma^{(c)}(\Omega) \right),
    \label{eq:lkl:flux}
\end{equation}
where $N^{(c)}$ (resp. $B^{(c)}$) is the observed (resp. background-expected) number of events in each category, and the product is performed over the set of four event categories ($c \in \mathcal{C}$, $\mathcal{C} =$ \{upgoing tracks, downgoing tracks, upgoing showers, downgoing showers\}). Given the selection optimisation presented in \autoref{sec:analysis}, the value of $B^{(c)}$ is fixed to $\num{2.7e-3}$, independently for all categories.

A Bayesian method is employed to obtain constraints on the neutrino flux normalisation $\phi$ starting from this likelihood. A flat prior on $\phi$ is employed, the systematic uncertainties described in \autoref{sec:syst} are encoded in Gaussian priors on $B^{(c)}$ and on $a_\gamma^{(c)}$ with the standard deviations corresponding to the uncertainties reported there, and the GW skymap $\mathcal{P}(\Omega)$ is used as a prior on $\Omega$.

The obtained posterior probability is then marginalised over all the nuisance parameters (background, acceptance, direction) by using MC integration techniques: toy samples (t.s.) are generated with values of $B^{(c)}$ and $a_\gamma^{(c)}$ following the priors, and the posterior samples from GW catalogues can be used directly for the sampling of $\Omega$. The marginalised posterior probability distribution is computed as
\begin{equation}
    P(\phi) = C \sum_{s \in \rm t.s.} \mathcal{L} \left( \{N^{(c)}\}; \{B^{(c)}\}_s, \{a_\gamma^{(c)}\}_s, \Omega_s, \phi, \gamma \right),
    \label{eq:post:flux}
\end{equation}
where $C$ is a normalisation constant that can be determined numerically by ensuring $\int_0^\infty P(\phi) {\rm d}\phi = 1$. The 90\% upper limit $\phi_{90}$ is finally obtained by solving $\int_0^{\phi_{90}} P(\phi) {\rm d}\phi = 0.90$.

\subsection{Constraints on the total energy}
\label{sec:etot}

Similarly to the incoming neutrino flux on Earth, one may also constrain the total energy $E_{\rm tot, \nu}$ emitted in neutrinos, correcting for the source distance. This can be done under a specific assumption on the spatial distribution of the neutrino emission around the source, e.g., either isotropic or collimated into a jet. One may also consider the ratio between the total energy emitted in neutrinos and the energy $E_{\rm GW}$ radiated in GW: $f_\nu = E_{\rm tot, \nu}/E_{\rm GW}$.

\paragraph{Isotropic emission.}

In the case of a source emitting isotropically, the total energy emitted in neutrinos can be computed as
\begin{equation}
    E_{\rm tot, \nu}^{\rm iso} = 4\pi D_{L}^2 \int_{E_{\min}}^{E_{\max}} E \times \dfrac{{\rm d}N}{{\rm d}E} \, {\rm d}E = \phi \times 4\pi D_{L}^2 \int_{E_{\min}}^{E_{\max}} E^{-\gamma+1} \, {\rm d}E,
    \label{eq:eiso}
\end{equation}
where $D_L$ is the source luminosity distance, and $E_{\min}$, $E_{\max}$ are the integration bounds. For this analysis, these bounds are fixed to $E_{\min} = \SI{5}{\giga\electronvolt}$ and $E_{\max} = \SI{e8}{\giga\electronvolt}$, which is the typical range where the emission is expected in most of the models (e.g., \cite{Ahlers:2019fwz}).

Since the total energy $E^{\rm iso}_{\rm tot, \nu}$ is proportional to the flux normalisation $\phi$, the likelihood from Equation \eqref{eq:lkl:flux} can be rewritten in terms of $E^{\rm iso}_{\rm tot, \nu}$ instead of $\phi$. The marginalised posterior probability and the upper limits are obtained similarly, where the luminosity distance is also extracted from GW posterior samples to be used in MC integration toy samples and a flat prior on $E^{\rm iso}_{\rm tot, \nu}$ is assumed. A similar rewriting is possible to obtain the limits on $f^{\rm iso}_\nu = E^{\rm iso}_{\rm tot, \nu}/E_{\rm GW}$. This is a relevant parameter if it is assumed that the total energy emitted in neutrinos scales with the GW emission.

\paragraph{Non-isotropic scenarios.}

One may also consider the total energy $E_{\rm tot, \nu}$ for a given non-isotropic model, and the corresponding likelihoods and posteriors can be written using the relevant GW parameters for the marginalisation. For non-isotropic emission, a simple von Mises~\cite{10.1098/rspa.1953.0064} jet model is considered:
\begin{equation}
    p(\theta; \omega) = \dfrac{1}{4\pi \omega^2 \sinh(1/\omega^2)} \times \exp(\cos\theta / \omega^2),
\end{equation}
where $\omega$ characterises the jet opening and $\theta$ is the angle with respect to the jet central direction. In the following, the jet is assumed to be collinear with the total angular momentum of the merger, such that $\theta = \theta_{\rm jn}$ from the GW data release. The total energy emitted in neutrinos for a power-law spectrum is then
\begin{equation}
    E_{\rm tot, \nu} = \phi \times 2\pi D_{L}^2 \int_{E_{\min}}^{E_{\max}} E^{-\gamma+1} \, {\rm d}E \times \dfrac{\int p(\theta; \omega) \sin\theta \, {\rm d}\theta}{p(\theta_{\rm jn}; \omega)},
\end{equation}
where the last term is the correction corresponding to the jet visibility from Earth (in the isotropic case, this term would be $2$, such that the Equation \eqref{eq:eiso} is retrieved). For a given jet opening $\omega$, the limits on $E_{\rm tot, \nu}$ or the corresponding $f_\nu$ may be derived as described in the isotropic case, including $\theta_{\rm jn}$ variable in MC integration toy samples.

\subsection{Stacking analysis}
\label{sec:stacking}

The GW catalogues published by LIGO/Virgo may contain populations of sources with similar neutrino emissions, which could be more efficiently constrained by performing stacking analyses. A first version of such an analysis is presented here with two categories of GW events: the 72 BBH mergers on one side and the 7 NSBH mergers on the other. While differing in its precise implementation in this study, the method follows the ideas initially presented in \cite{Veske:2020yjt} and already applied in \cite{Super-Kamiokande:2021dav} with Super-Kamiokande data.

The stacking approach assumes that all objects in the selected population have the same emission, either in terms of total energy $E_{\rm tot, \nu}$ or of $f_\nu$. In the non-isotropic scenarios, GW sources in a given population may have different jet inclinations but the shape of the jet (the parameter $\omega$ for the von Mises model) is considered to be the same for all the objects.

Assuming a flat prior on the signal parameters and considering all observations as independent, the posterior distribution for a given population $\mathcal{S}$ may be written as
\begin{equation}
    P_{\mathcal{S}}(X; Y) = C \prod_{i \in \mathcal{S}} P_i(X; Y, Z_i),
\end{equation}
where $C$ is a normalisation constant, $X$ is the signal parameter to be constrained ($E_{\rm tot, \nu}$ or $f_\nu$), $Y$ may denote potential common parameters (such as $\omega$), and $Z_i$ represents the parameters which may differ from one GW event to another within $\mathcal{S}$. Upper limits on $X$ can then be obtained with the same method as before.

The stacking approach is particularly interesting to constrain the non-isotropic models, as is detailed in the next section.

\section{Results}
\label{sec:results}

A total of 80 GW events out of the 83 observed during O3, including all candidates involving at least one neutron star, can be associated with exploitable ANTARES data.

For each follow-up, the neutrino event selection is optimised according to the procedure described in \autoref{sec:analysis} and the final number of selected events in time and spatial coincidence with the GW event in each category is extracted. No event has been selected for any of the follow-ups, which is fully compatible with the expected background $\sum_{i \in {\rm GWs}} \sum_{c \in \mathcal{C}} B_i^{(c)} \sim 0.82$. Therefore, only upper limits on the neutrino flux and other related quantities are reported in the following.

\autoref{tab:results:gwtc2} displays the 90\% upper limits on the incoming all-flavour neutrino emission assuming an $E^{-2}$ spectrum, along with GW information, for the events initially reported in the GWTC-2 catalogue. Similarly, the results for the events in GWTC-2.1 and GWTC-3 are presented in \autoref{tab:results:gwtc2.1} and \autoref{tab:results:gwtc3}, respectively. The all-flavour flux limits assuming an $E^{-2}$ spectrum stands mostly between $4$ and \SI{60}{\giga\electronvolt\per\square\centi\meter}.

\begin{table}[ht]
\caption{Summary of GWTC-2 follow-up results. The first four columns summarise the relevant GW information~\cite{LIGOScientific:2020ibl}, including the most probable merger type, the estimated median luminosity distance $D_L$, and the size of the region $\mathcal{R}_{90}$ containing 90\% of the localisation probability, while the three last ones show the 90\% upper limits on the all-flavour neutrino emission assuming an $E^{-2}$ spectrum (this work), in terms of the flux normalisation $E^2 {\rm d}N/{\rm d}E$, the total isotropic energy $E^{\rm iso}_{\rm tot, \nu}$, and the ratio $f_\nu^{\rm iso}$.}
\label{tab:results:gwtc2}
\centering
\begin{tabular}{|l|cS[table-format=6]S[table-format=5]|*{3}{S[table-format=3.1e2,table-auto-round,table-number-alignment=center,retain-zero-exponent]}|}
\hline
{GW name} & {Type} & {$D_L$} & {$\mathcal{R}_{90}$ area} & \multicolumn{3}{c|}{Upper limits on neutrino emission} \\
{} & {} & {} & {} & {$E^2 {\rm d}N/{\rm d}E$} & {$E^{\rm iso}_{\rm tot, \nu}$} & {$f_{\nu}^{\rm iso}$} \\
{} & {} & {\si{\mega\parsec}} & {\si{\deg\squared}} & {\si{\giga\electronvolt\per\square\centi\meter}} & {\si{\erg}} & {} \\
\hline
GW190412 & BBH & 734 & 24 & 1.75e+01 & 3.40e+55 & 1.67e+01 \\
GW190413\_052954 & BBH & 4190 & 1383 & 6.37e+01 & 1.80e+57 & 3.67e+02 \\
GW190413\_134308 & BBH & 5182 & 520 & 5.35e+00 & 5.90e+56 & 9.06e+01 \\
GW190421\_213856 & BBH & 3166 & 1023 & 1.04e+01 & 3.35e+56 & 6.12e+01 \\
GW190424\_180648 & BBH & 2568 & 25902 & 4.62e+01 & 1.23e+57 & 2.02e+02 \\
GW190425 & BNS & 157 & 9881 & 2.13e+01 & 2.31e+54 & 8.56e+00 \\
GW190426\_152155 & NSBH & 377 & 1392 & 2.61e+01 & 2.44e+55 & 8.69e+01 \\
GW190503\_185404 & BBH & 1527 & 97 & 4.44e+00 & 4.46e+55 & 7.68e+00 \\
GW190512\_180714 & BBH & 1462 & 229 & 4.62e+00 & 3.99e+55 & 1.63e+01 \\
GW190513\_205428 & BBH & 2190 & 494 & 6.29e+00 & 9.01e+55 & 2.28e+01 \\
GW190514\_065416 & BBH & 4988 & 2402 & 1.14e+01 & 1.12e+57 & 2.40e+02 \\
GW190517\_055101 & BBH & 2270 & 468 & 1.75e+01 & 6.73e+56 & 9.47e+01 \\
GW190519\_153544 & BBH & 3023 & 770 & 1.22e+01 & 7.23e+56 & 7.28e+01 \\
GW190521 & BBH & 4567 & 937 & 1.19e+01 & 9.64e+56 & 7.34e+01 \\
GW190521\_074359 & BBH & 1244 & 509 & 1.75e+01 & 9.87e+55 & 1.65e+01 \\
GW190527\_092055 & BBH & 3563 & 3795 & 4.11e+01 & 4.56e+57 & 6.38e+02 \\
GW190602\_175927 & BBH & 3138 & 721 & 5.40e+00 & 2.42e+56 & 2.60e+01 \\
GW190620\_030421 & BBH & 3211 & 6674 & 2.86e+01 & 1.40e+57 & 1.57e+02 \\
GW190630\_185205 & BBH & 956 & 1275 & 2.65e+01 & 9.92e+55 & 2.23e+01 \\
GW190701\_203306 & BBH & 2152 & 47 & 6.07e+00 & 9.75e+55 & 1.32e+01 \\
GW190706\_222641 & BBH & 5184 & 611 & 9.72e+00 & 1.25e+57 & 1.26e+02 \\
GW190707\_093326 & BBH & 791 & 1343 & 2.42e+01 & 7.13e+55 & 4.50e+01 \\
GW190708\_232457 & BBH & 888 & 13701 & 2.60e+01 & 7.59e+55 & 3.17e+01 \\
GW190719\_215514 & BBH & 4786 & 2281 & 5.19e+03 & 2.14e+59 & 3.01e+04 \\
GW190720\_000836 & BBH & 906 & 517 & 2.52e+01 & 5.98e+55 & 3.62e+01 \\
GW190727\_060333 & BBH & 3609 & 861 & 1.58e+01 & 8.74e+56 & 1.41e+02 \\
GW190731\_140936 & BBH & 4034 & 3042 & 1.84e+01 & 1.43e+57 & 2.51e+02 \\
GW190803\_022701 & BBH & 3750 & 1538 & 2.20e+01 & 9.74e+56 & 1.93e+02 \\
GW190814 & NSBH & 241 & 20 & 4.60e+00 & 9.49e+53 & 2.21e+00 \\
GW190828\_063405 & BBH & 2160 & 525 & 8.73e+00 & 1.35e+56 & 2.61e+01 \\
GW190828\_065509 & BBH & 1658 & 637 & 7.12e+00 & 9.57e+55 & 4.52e+01 \\
GW190909\_114149 & BBH & 4924 & 4071 & 3.04e+01 & 3.74e+57 & 7.08e+02 \\
GW190910\_112807 & BBH & 1670 & 10014 & 2.18e+01 & 3.81e+56 & 6.25e+01 \\
GW190915\_235702 & BBH & 1715 & 386 & 5.37e+00 & 7.21e+55 & 1.54e+01 \\
GW190924\_021846 & BBH & 572 & 379 & 8.89e+00 & 1.05e+55 & 1.13e+01 \\
GW190929\_012149 & BBH & 3901 & 1846 & 1.44e+01 & 1.44e+57 & 2.25e+02 \\
GW190930\_133541 & BBH & 786 & 1829 & 1.54e+01 & 3.51e+55 & 2.40e+01 \\
\hline
\end{tabular}
\end{table}

\begin{table}[ht]
\caption{Same as \autoref{tab:results:gwtc2} for the additional events in the GWTC-2.1 catalogue~\cite{LIGOScientific:2021usb}.}
\label{tab:results:gwtc2.1}
\centering
\begin{tabular}{|l|cS[table-format=6]S[table-format=5]|*{3}{S[table-format=3.1e2,table-auto-round,table-number-alignment=center,retain-zero-exponent]}|}
\hline
GW name & Type & {Distance} & {$\mathcal{R}_{90}$ area} & \multicolumn{3}{c|}{Upper limits on neutrino emission} \\
{} & {} & {} & {} & {$E^2 {\rm d}N/{\rm d}E$} & {$E^{\rm iso}_{\rm tot, \nu}$} & {$f_{\nu}^{\rm iso}$} \\
{} & {} & {\si{\mega\parsec}} & {\si{\deg\squared}} & {\si{\giga\electronvolt\per\square\centi\meter}} & {\si{\erg}} & {} \\
\hline
GW190403\_051519 & BBH & 10853 & 4250 & 2.89e+01 & 1.67e+58 & 1.82e+03 \\
GW190426\_190642 & BBH & 5995 & 8030 & 3.61e+01 & 5.42e+57 & 3.51e+02 \\
GW190725\_174728 & BBH & 1128 & 2435 & 2.52e+01 & 1.46e+56 & 1.28e+02 \\
GW190805\_211137 & BBH & 7235 & 3089 & 4.12e+01 & 9.94e+57 & 1.47e+03 \\
GW190916\_200658 & BBH & 6180 & 3573 & 3.35e+01 & 6.95e+57 & 1.16e+03 \\
GW190917\_114630 & NSBH & 741 & 1801 & 1.20e+01 & 2.80e+55 & 7.79e+01 \\
GW190925\_232845 & BBH & 973 & 955 & 6.72e+00 & 2.38e+55 & 7.68e+00 \\
GW190926\_050336 & BBH & 5073 & 2212 & 5.59e+01 & 9.60e+57 & 2.01e+03 \\
\hline
\end{tabular}
\end{table}

\begin{table}[ht]
\caption{Same as \autoref{tab:results:gwtc2} for the events in the GWTC-3 catalogue~\cite{LIGOScientific:2021djp}.}
\label{tab:results:gwtc3}
\centering
\begin{tabular}{|l|cS[table-format=6]S[table-format=5]|*{3}{S[table-format=3.1e2,table-auto-round,table-number-alignment=center,retain-zero-exponent]}|}
\hline
GW name & Type & {Distance} & {$\mathcal{R}_{90}$ area} & \multicolumn{3}{c|}{Upper limits on neutrino emission} \\
{} & {} & {} & {} & {$E^2 {\rm d}N/{\rm d}E$} & {$E^{\rm iso}_{\rm tot, \nu}$} & {$f_{\nu}^{\rm iso}$} \\
{} & {} & {\si{\mega\parsec}} & {\si{\deg\squared}} & {\si{\giga\electronvolt\per\square\centi\meter}} & {\si{\erg}} & {} \\
\hline
GW191103\_012549 & BBH & 927 & 2520 & 4.34e+01 & 1.61e+56 & 9.95e+01 \\
GW191105\_143521 & BBH & 1141 & 730 & 1.45e+02 & 6.31e+56 & 4.31e+02 \\
GW191109\_010717 & BBH & 1386 & 1784 & 1.24e+01 & 1.21e+56 & 1.76e+01 \\
GW191113\_071753 & BBH & 1469 & 2993 & 3.10e+01 & 3.72e+56 & 2.95e+02 \\
GW191126\_115259 & BBH & 1618 & 1514 & 1.06e+02 & 7.95e+56 & 4.24e+02 \\
GW191127\_050227 & BBH & 3614 & 1499 & 2.07e+01 & 1.76e+57 & 2.50e+02 \\
GW191129\_134029 & BBH & 774 & 851 & 1.48e+01 & 3.58e+55 & 2.68e+01 \\
GW191204\_110529 & BBH & 1975 & 4747 & 4.57e+01 & 8.94e+56 & 2.81e+02 \\
GW191204\_171526 & BBH & 637 & 365 & 8.63e+00 & 1.17e+55 & 6.77e+00 \\
GW191215\_223052 & BBH & 1937 & 603 & 1.30e+01 & 2.50e+56 & 7.75e+01 \\
GW191216\_213338 & BBH & 339 & 488 & 1.13e+01 & 5.34e+54 & 3.51e+00 \\
GW191219\_163120 & NSBH & 569 & 2232 & 1.74e+01 & 2.78e+55 & 1.90e+02 \\
GW191222\_033537 & BBH & 2991 & 2299 & 5.62e+01 & 2.69e+57 & 4.60e+02 \\
GW191230\_180458 & BBH & 4296 & 1012 & 1.15e+01 & 9.18e+56 & 1.40e+02 \\
GW200105\_162426 & NSBH & 266 & 7882 & 1.75e+01 & 6.11e+54 & 1.68e+01 \\
GW200112\_155838 & BBH & 1248 & 4250 & 1.81e+01 & 1.36e+56 & 2.57e+01 \\
GW200115\_042309 & NSBH & 298 & 519 & 6.84e+00 & 2.81e+54 & 1.17e+01 \\
GW200128\_022011 & BBH & 3405 & 2677 & 1.83e+02 & 7.57e+57 & 1.35e+03 \\
GW200129\_065458 & BBH & 883 & 87 & 8.80e+00 & 2.82e+55 & 5.38e+00 \\
GW200202\_154313 & BBH & 411 & 160 & 5.37e+00 & 3.24e+54 & 2.40e+00 \\
GW200208\_130117 & BBH & 2258 & 39 & 7.22e+00 & 1.51e+56 & 2.99e+01 \\
GW200208\_222617 & BBH & 4547 & 1889 & 1.57e+01 & 2.52e+57 & 4.90e+02 \\
GW200209\_085452 & BBH & 3447 & 925 & 2.50e+02 & 2.24e+57 & 4.53e+02 \\
GW200210\_092255 & NSBH & 959 & 1830 & 2.28e+01 & 8.57e+55 & 1.87e+02 \\
GW200216\_220804 & BBH & 3997 & 3010 & 7.80e+00 & 6.45e+56 & 1.03e+02 \\
GW200219\_094415 & BBH & 3447 & 702 & 6.41e+00 & 3.74e+56 & 8.05e+01 \\
GW200220\_061928 & BBH & 6321 & 3485 & 1.87e+01 & 5.89e+57 & 6.59e+02 \\
GW200220\_124850 & BBH & 4172 & 3169 & 4.53e+01 & 4.26e+57 & 9.34e+02 \\
GW200224\_222234 & BBH & 1677 & 51 & 1.32e+01 & 1.48e+56 & 2.37e+01 \\
GW200225\_060421 & BBH & 1144 & 516 & 2.15e+01 & 1.05e+56 & 4.26e+01 \\
GW200302\_015811 & BBH & 1545 & 7011 & 2.77e+01 & 3.06e+56 & 7.44e+01 \\
GW200306\_093714 & BBH & 2228 & 4371 & 1.08e+01 & 2.66e+56 & 7.37e+01 \\
GW200308\_173609 & BBH & 8867 & 18705 & 3.50e+01 & 3.27e+58 & 1.75e+04 \\
GW200311\_115853 & BBH & 1152 & 37 & 1.74e+01 & 8.65e+55 & 1.76e+01 \\
GW200322\_091133 & BBH & 8302 & 31571 & 3.64e+01 & 2.98e+58 & 4.88e+04 \\
\hline
\end{tabular}
\end{table}

\begin{figure*}[hbtp]
    \subfloat[]{\includegraphics[width=0.48\textwidth]{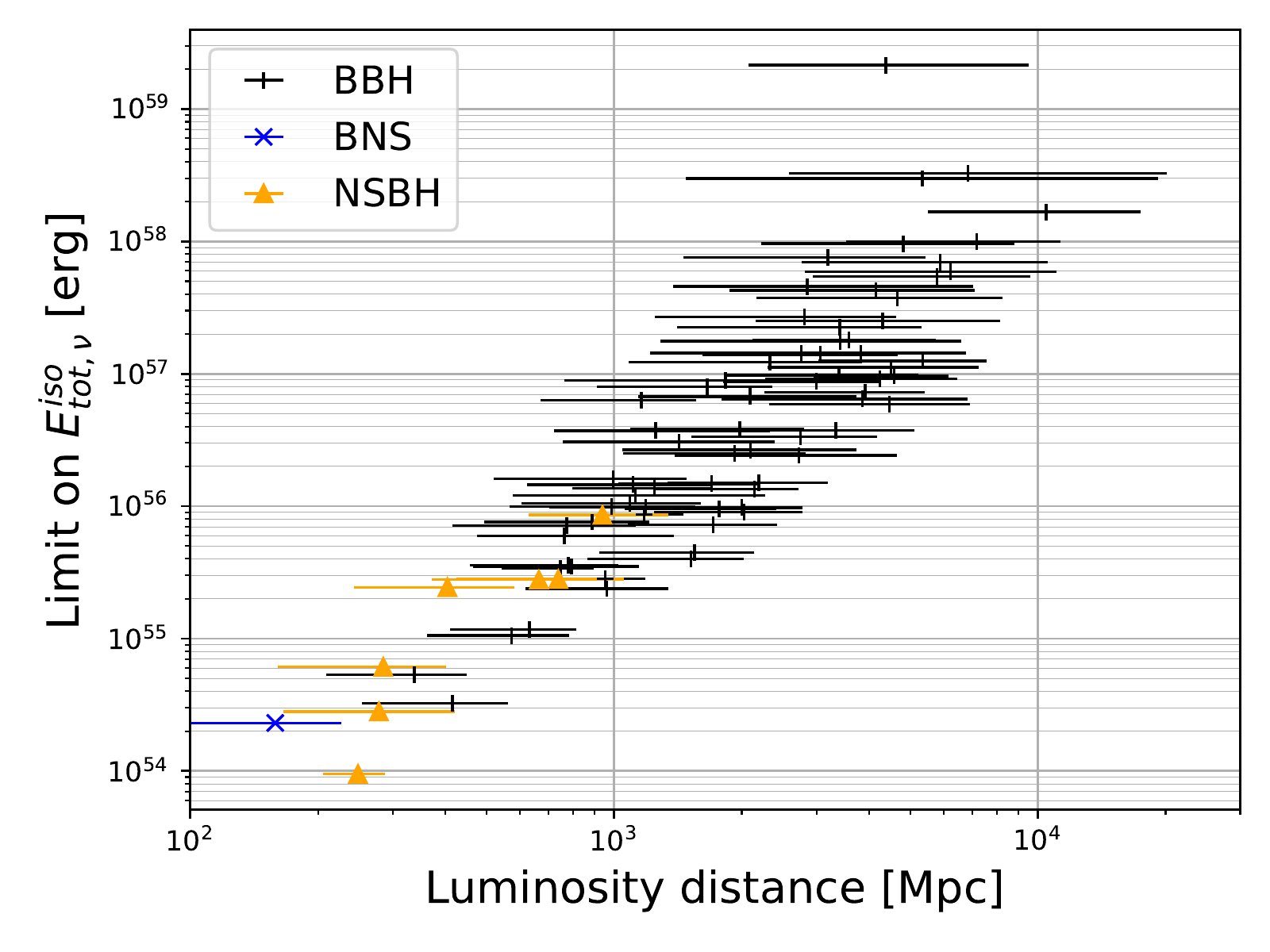}}
    \subfloat[]{\includegraphics[width=0.48\textwidth]{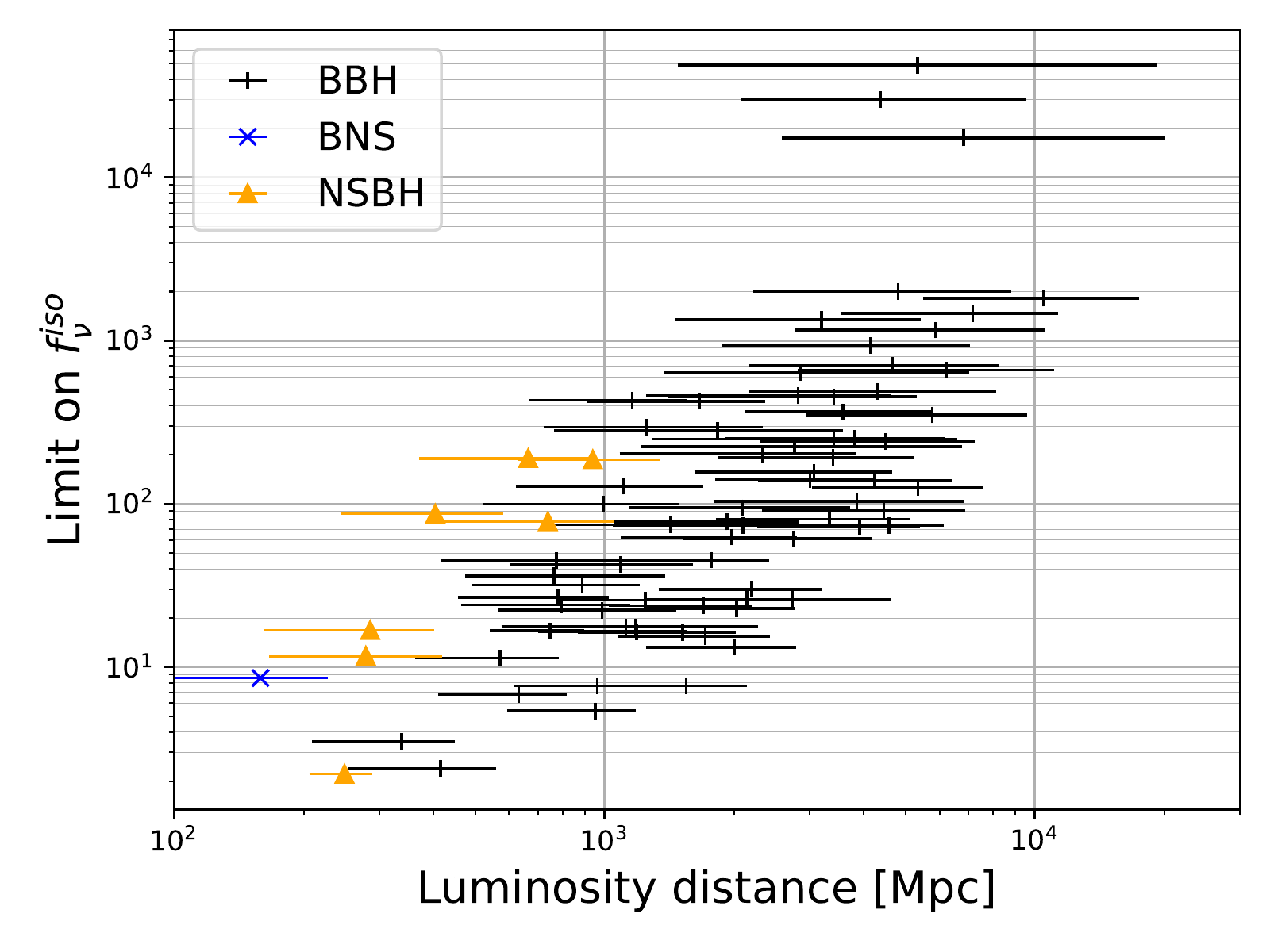}}
    \caption{90\% upper limits on the total energy $E^{\rm iso}_{\rm tot, \nu}$ emitted in neutrinos of all flavours (left) and on $f^{\rm iso}_\nu = E^{\rm iso}_{\rm tot, \nu}/E_{\rm GW}$ (right) as a function of the source luminosity distance, assuming an $E^{-2}$ spectrum and isotropic emission. The horizontal bars indicate the $5-95\%$ range of the luminosity distance estimate, and the markers/colours correspond to the different source categories.}
    \label{fig:limits:eiso}
\end{figure*}

\autoref{fig:limits:eiso} shows the limits on the total energy emitted in all-flavour neutrinos and on $f^{\rm iso}_\nu = E^{\rm iso}_{\rm tot, \nu}/E_{\rm GW}$, assuming isotropic emission, for events from the three catalogues. The limits range from \num{e54} to \SI{e59}{\erg} and are mostly following the expected $D_L^2$ trend.

\begin{figure*}[hbtp]
    \subfloat[]{\includegraphics[width=0.48\textwidth]{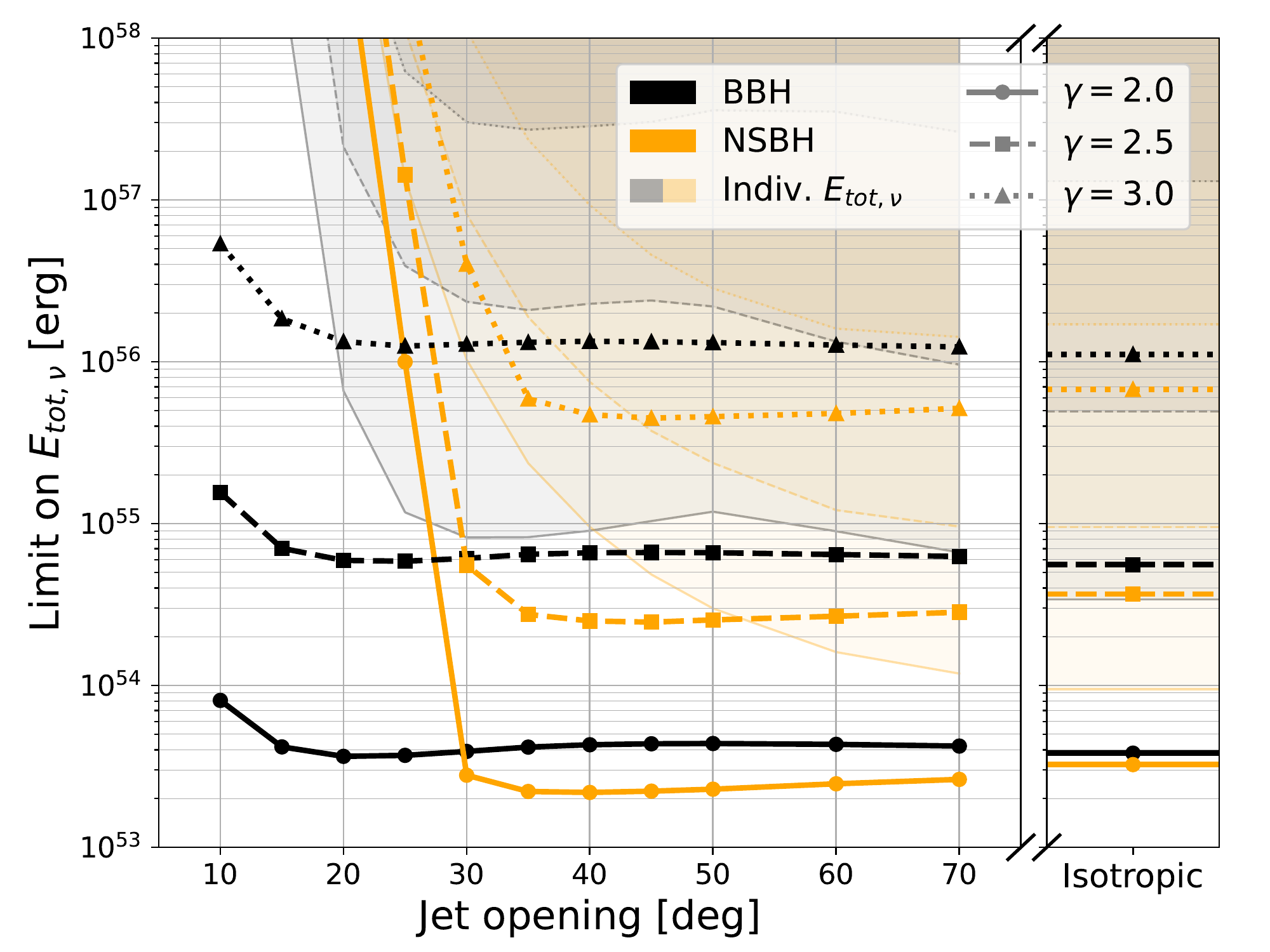}}
    \subfloat[]{\includegraphics[width=0.48\textwidth]{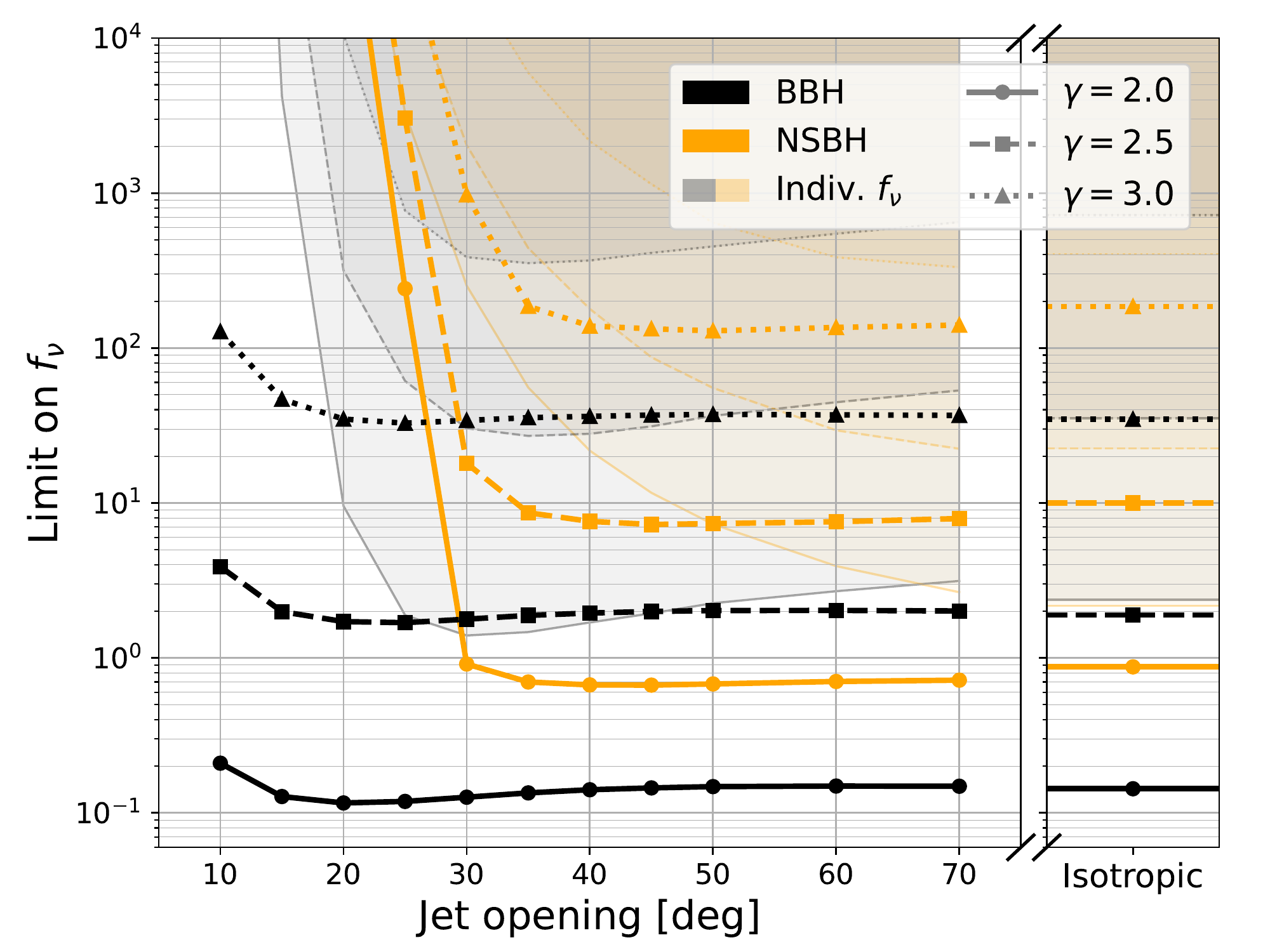}}
    \caption{90\% upper limits on the per-source typical total energy emitted in neutrinos (left) and on $f_\nu$ (right) as a function of the jet opening angle (for von Mises model) and for the isotropic case (rightmost entry). The black (orange) curves correspond to the BBH (NSBH) stacking of mergers detected during O3. The different markers and line styles correspond to different spectral indices $\gamma$. The filled areas with the corresponding colours (merger category) and line styles (spectral index) indicate the envelope of the best individual upper limits.}
    \label{fig:limits:stacking}
\end{figure*}

Population studies are performed by stacking (a) all 72 BBH events, and (b) the 7 NSBH candidates. \autoref{fig:limits:stacking} shows the stacked upper limits on the typical total energy emitted in neutrinos (or the ratio $f_\nu$) within these two categories, considering jetted emission with the von Mises model (as a function of the jet opening $\omega$) as well as for isotropic emission, and assuming $E^{-\gamma}$ spectra with $\gamma = 2.0,\,2.5,\,3.0$. 

For individual follow-ups, as $\theta_{\rm jn}$ is not strongly constrained by GW detections, von Mises emission constraints are mainly dominated by geometrical effects: when the jet opening angle is narrow, there is a very small chance that the Earth is aligned with the jet, while a wide opening leads to a larger spread in total energy, which is hence more difficult to constrain. The limits on the total energy as a function of the jet opening are thus expected to show a typical parabola shape corresponding to a trade-off between these two effects, as illustrated by the envelope of individual limits shown in \autoref{fig:limits:stacking}. While the individual results for jetted emission are of little interest due to these limitations, the stacking analysis of this scenario yields some benefit as, statistically, some of the sources are expected to point toward Earth for large enough jet opening angles. This can be seen on \autoref{fig:limits:stacking}, in particular for the BBH event sample, with the largest statistics, where the neutrino emission can be constrained down to opening angles as small as $\omega \sim 10-\SI{30}{\degree}$.

For the BBH stacking, the stacked limits in the nominal scenario ($\gamma = 2$ and isotropic emission) are $E^{\rm iso}_{\rm tot, \nu} < \SI{3.8e53}{\erg}$ and $f_\nu < 0.14$, while the best limits from individual follow-ups (for GW200202\_154313) are respectively $\SI{3.2e54}{\erg}$ and $2.4$. For the NSBH population, the corresponding limits are $E^{\rm iso}_{\rm tot, \nu} < \SI{3.2e53}{\erg}$ and $f_\nu < 0.88$, with the best individual limits being (for GW190814) $\SI{9.5e53}{\erg}$ and $2.2$, respectively.

Finally, it is worth noticing that the statistical power of the BBH stacking already allows the exploration of reasonable values for $f_\nu$ (in the context of the assumed simple power-law spectrum), while all individual limits on this parameter are above 1, corresponding to a neutrino emission higher than the GW one.

Supplemental material with all numerical results for individual follow-ups and from the stacking analysis, for the various jet scenarios and assumed spectral indices, can be provided under request.

\section{Discussion and conclusions}
\label{sec:concl}

The search for neutrino counterparts to GW signals detected during the O3 run with the ANTARES detector yields no significant excess. This null result is used to extract upper limits on the neutrino emission, both in terms of the flux normalisation and of the total energy emitted in neutrinos, $E^{\rm iso}_{\rm tot, \nu}$, in particular for the standard case with an $E^{-2}$ spectrum and isotropic emission.

The ANTARES analysis presented in this article benefits from the different event topologies in the detector, leading to all-flavour limits for neutrino energies above \SI{100}{\giga\electronvolt}. The constraints are also interpreted in the context of population studies, for which limits can be put on the typical neutrino emission of different categories of sources (here, BBH and NSBH). As the most recent astrophysical models (e.g., \cite{Kimura:2018vvz,Ahlers:2019fwz,Decoene:2019eux}) seem to be out of reach for current detectors, such stacking studies may be one of the most promising approaches to identify joint emitters of neutrinos and GWs. More refined stackings may be studied by considering e.g., sub-populations among BBH mergers, such as the ones having similar source characteristics or environments.

The Bayesian analysis performed here considers flat priors, such that the stacking analysis is performed easily by multiplying the individual posterior distributions. Different priors (e.g., non-informative Jeffreys prior~\cite{Jeffreys}) introduce correlations between the GW follow-ups and would only be possible by performing the posterior marginalisation over all GW events at once, therefore requiring more sophisticated integration techniques than the one employed in this paper.

First constraints on jetted emission for a von Mises model and for different spectral indices are derived as well. In the absence of the determination of any jet direction from the GW data or any electromagnetic observations, these results are only indirectly deducted from the stacking analysis. In future observation campaigns, any such measurement, even for one or a few sources, would allow for deriving direct constraints, as well as considering more complex jet models.

Though the ANTARES detector has been decommissioned in 2022, the KM3NeT experiment \cite{KM3Net:2016zxf} has taken over the observation of the neutrino sky from the depths of the Mediterranean Sea. Data from the first few lines of KM3NeT/ORCA has already been used to perform an initial search for neutrino counterparts to the O3 events \citep{KM3NeT:GW}. With more lines being deployed up to $\sim$ 2026, KM3NeT/ARCA and ORCA are expected to provide complementary coverage of the transient sky with respect to IceCube, increasing the chances to get significant individual detections and creating new opportunities for joint analyses.

The next GW observing period, O4, is expected to start in 2023, including the KAGRA detector in addition to LIGO and Virgo \citep{KAGRA:2013rdx}. With an expected number of detections  of the order of one hundred per year, population studies will become even more relevant. 

Lastly, combinations with ZTF \citep{Bellm_2018} or Vera C. Rubin \citep{Andreoni:2021epw} surveys and Target of Opportunity programs may help pinpoint candidate host galaxies for observed binary mergers, and allow more detailed analyses using the electromagnetic information and the environment characterisation as inputs for the prediction and constraint of the neutrino emission. More generally, there are still only a few models on the processes leading to neutrino production during binary merger events in the literature~\cite{Kimura:2018vvz,Kimura:2017kan,Kotera:2016dmp,Ahlers:2019fwz,Decoene:2019eux}, and further theoretical developments in the coming years may bring new light to past, present, and future observations.

\acknowledgments{The authors acknowledge the financial support of the funding agencies:
Centre National de la Recherche Scientifique (CNRS), Commissariat \`a
l'\'ener\-gie atomique et aux \'energies alternatives (CEA),
Commission Europ\'eenne (FEDER fund and Marie Curie Program),
LabEx UnivEarthS (ANR-10-LABX-0023 and ANR-18-IDEX-0001),
R\'egion Alsace (contrat CPER), R\'egion Provence-Alpes-C\^ote d'Azur,
D\'e\-par\-tement du Var and Ville de La
Seyne-sur-Mer, France;
Bundesministerium f\"ur Bildung und Forschung
(BMBF), Germany; 
Istituto Nazionale di Fisica Nucleare (INFN), the European Union’s Horizon 2020 research and innovation programme under the Marie Sklodowska-Curie grant agreement No 754496, Italy;
Nederlandse organisatie voor Wetenschappelijk Onderzoek (NWO), the Netherlands;
Executive Unit for Financing Higher Education, Research, Development and Innovation (UEFISCDI), Romania;
Grants PID2021-124591NB-C41, -C42, -C43 funded by MCIN/AEI/ 10.13039/501100011033 and, as appropriate, by “ERDF A way of making Europe”, by the “European Union” or by the “European Union NextGenerationEU/PRTR”,  Programa de Planes Complementarios I+D+I (refs. ASFAE/2022/023, ASFAE/2022/014), Programa Prometeo (PROMETEO/2020/019) and GenT (refs. CIDEGENT/2018/034, /2019/043, /2020/049. /2021/23) of the Generalitat Valenciana, Junta de Andaluc\'{i}a (ref. P18-FR-5057), EU: MSC program (ref. 101025085), Programa Mar\'{i}a Zambrano (Spanish Ministry of Universities, funded by the European Union, NextGenerationEU), Spain;
Ministry of Higher Education, Scientific Research and Innovation, Morocco, and the Arab Fund for Economic and Social Development, Kuwait.
We also acknowledge the technical support of Ifremer, AIM and Foselev Marine
for the sea operation and the CC-IN2P3 for the computing facilities.}

\bibliographystyle{JHEP}
\bibliography{references}

\end{document}